\documentclass[10pt,sigconf,letterpaper, nonacm]{acmart}
\usepackage[utf8]{inputenc}

\usepackage[T1]{fontenc}

\usepackage{microtype}

\usepackage{bm,mathtools,color}

\usepackage[normalem]{ulem}
\usepackage{enumitem}

\usepackage{booktabs}
\usepackage{multirow}
\usepackage{subcaption}
\usepackage{tabularx}
\usepackage{multicol}
\usepackage{multirow}
\usepackage{colortbl}
\usepackage{hhline}
\usepackage{stfloats}
\usepackage{float}

\newcolumntype{Y}{>{\centering\arraybackslash}X}

\usepackage{booktabs}
\usepackage{multirow}
\usepackage{graphicx}

\usepackage{pifont}
\usepackage{amsmath}
\usepackage{ifthen}
\usepackage{dsfont}

\usepackage{listings}
\usepackage{xcolor}
\usepackage{ltablex}
\usepackage{longtable}
\usepackage{multicol}

\usepackage{tikz}
\usetikzlibrary{bayesnet}
\usepackage{subfloat}
\usepackage[linguistics]{forest}

\usepackage{algorithm}
\usepackage{algpseudocode}

\algrenewcommand\algorithmicrequire{\textbf{Input:}}
\algrenewcommand\algorithmicensure{\textbf{Output:}}
\algnewcommand{\LeftComment}[1]{\Statex \(\triangleright\) #1}

\usepackage{caption}

\usepackage{cancel}

\usepackage{xspace}

\usepackage[leftmargin=0em]{quoting}
\usepackage{soul}
\usepackage[most]{tcolorbox}

\newcommand{\ensuretext}[1]{#1}

\newcommand{\marker}[2]{\ensuremath{^{\textsc{#1}}_{\textsc{#2}}}}
\newcommand{\authorcomment}[4]{\draftonly{\ensuretext{\textcolor{#3}{[\marker{#1}{#2} #4]}}}}

\newcommand{\joe}[1]{\authorcomment{J}{Z}{teal!80}{#1}}
\newcommand{\samuel}[1]{\authorcomment{S}{L}{purple!80}{#1}}
\newcommand{\minlan}[1]{\authorcomment{M}{Y}{cyan!80}{#1}}

\newcommand{\bfsection}[1]{\noindent\textbf{#1.}}

\newcommand{\presec}{\vspace{-0.03in}}%{\vspace{-0.12in}}
\newcommand{\postsec}{\vspace{-0.04in}}%{\vspace{-0.05in}}
\newcommand{\presub}{\vspace{-0.03in}}%{\vspace{-0.12in}}
\newcommand{\postsub}{\vspace{-0.03in}}%{\vspace{-0.05in}}

\newcommand{\prefigcaption}{\vspace{-0.05in}}%{\vspace{-0.08in}}
\newcommand{\postfigcaption}{\vspace{-0.2in}}%{\vspace{-0.15in}}
\newcommand{\posttablecaption}{\vspace{-0.1in}}%{\vspace{-0.15in}}

\newcommand{\draftonly}[1]{#1}
\renewcommand{\draftonly}[1]{}

\AtBeginDocument{%
  }

\setcopyright{acmlicensed}
\acmConference[Conference acronym 'XX]{Make sure to enter the correct
  conference title from your rights confirmation emai}{June 03--05,
  2018}{Woodstock, NY}
\acmISBN{978-1-4503-XXXX-X/18/06}

\begin{document}

\title{An LLM-based Agentic Framework for Accessible Network Control}
% \title{Talk to Your Networks with Natural Language: Towards Readily Accessible Network Control}

%% used to denote shared contribution to the research

\author{Samuel Lin}
\authornote{\small Work done while the author was a student at Harvard University.}
\affiliation{%
  \institution{Meta}
  \country{}
}
\email{samuellin@meta.com}

\author{Jiawei Zhou}
\affiliation{%
  \institution{Stony Brook University}
  \country{}
}
\email{jiawei.zhou.1@stonybrook.edu}

\author{Minlan Yu}
\affiliation{%
  \institution{Harvard University}
  \country{}
}
\email{minlanyu@g.harvard.edu}

\renewcommand{\shortauthors}{Lin et al.}

\begin{abstract}
Traditional approaches to network management have been accessible only to a handful of highly-trained network operators with significant expert knowledge. This creates barriers for lay users to easily manage their networks without resorting to experts. With recent development of powerful large language models (LLMs) for language comprehension, we design a system to make network management accessible to a broader audience of non-experts by allowing users to converse with networks in natural language. To effectively leverage advancements in LLMs, we propose an agentic framework that uses an intermediate representation to streamline configuration across diverse vendor equipment, retrieves the network state from memory in real-time, and provides an interface for external feedback.  We also conduct pilot studies to collect real user data of natural language utterances for network control, and present a visualization interface to facilitate dialogue-driven user interaction and enable large-scale data collection for future development. Preliminary experiments validate the effectiveness of our proposed system components with LLM integration on both synthetic and real user utterances. Through our data collection and visualization efforts, we pave the way for more effective use of LLMs and democratize network control for everyday users.\footnote{\small Data and code available at \url{https://github.com/jzhou316/LLM-networking-control}.}

\end{abstract}

\maketitle

\section{Introduction}
\label{sec-intro}

\joe{add references} Network management is a challenging task. As networks grow in size and complexity, managing them becomes labor-intensive and error-prone even for experienced network operators \cite{aditya-akella-network-management-difficult, robotron}. Moreover, because modern networks rely on equipment from multiple vendors, network management remains accessible only to a limited number of experts who are familiar with the specific tools, documentation, and configuration protocols unique to each vendor. This creates a steep learning curve for non-experts, which can stifle innovation and limit the democratization of network technologies.

\begin{figure}[t]
    \centering
    % \prefig
    \includegraphics[width=0.45\textwidth]{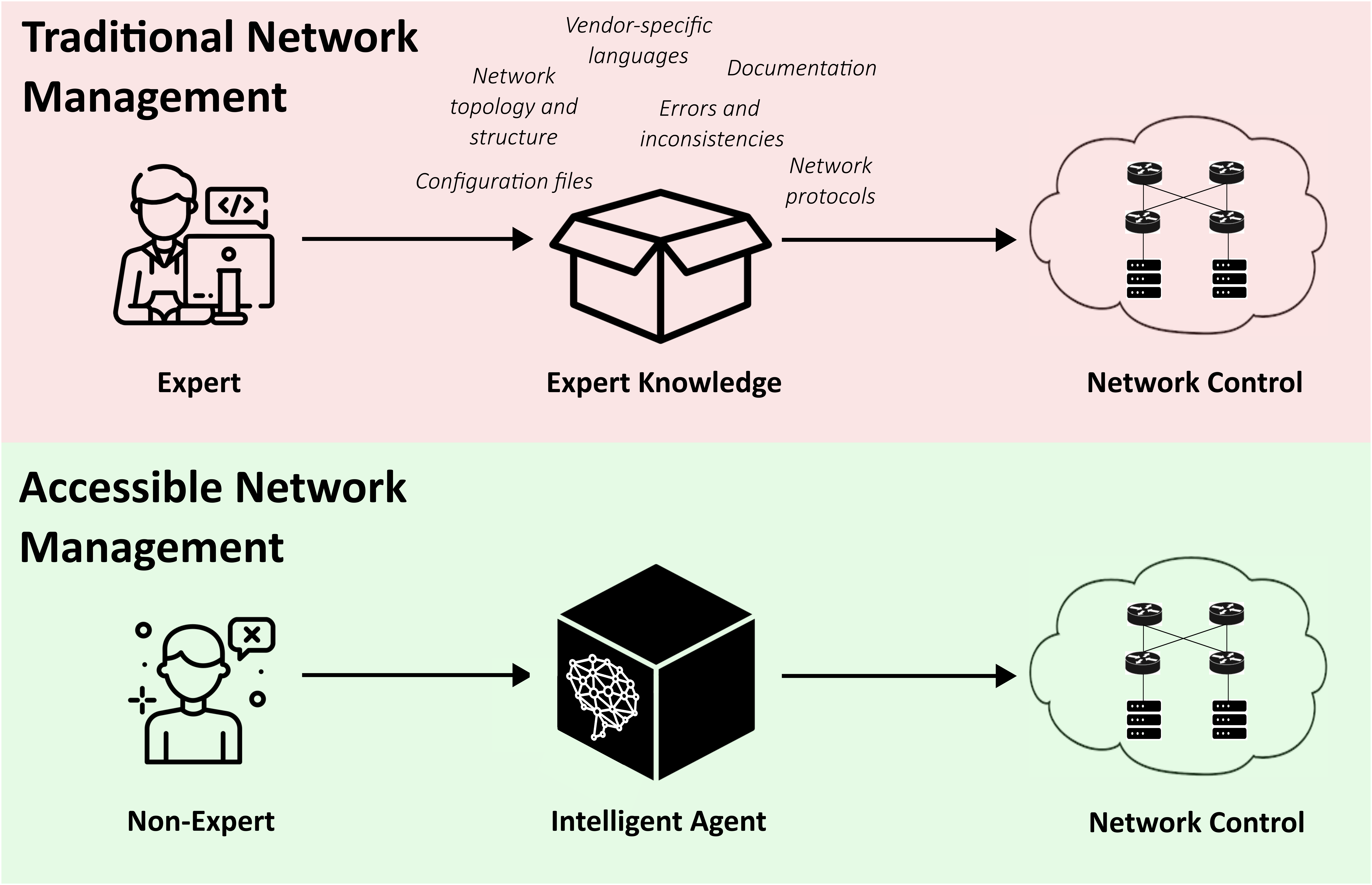}
    \prefigcaption\caption{A desirable vision for future network control.
    % \joe{change "Black-Box LLM" to "Black-Box System", to cite the figure before mentioning LLM}
    }\postfigcaption
    \label{fig:vision}
\end{figure}

\joe{can add a figure: previous vs. future network config. \\
Previous: experts, transparent system components (whith box), network confg; \\
current: non-experts, NL utterance, black box system, network config very easy.}\joe{added}
% Our vision is to make network control more accessible to non-experts across a wide range of network applications and scenarios.
We envision a future where network control is more accessible to non-experts across a wide range of network applications, as depicted by Figure~\ref{fig:vision}. For example, entry-level developers and end-users may lack the specialized  knowledge necessary to configure complex home or business networks. 
% For example, non-experts such as parents, elderly individuals, and small business owners often need to configure complex home or business networks without specialized technical knowledge.} 
\samuel{hotnets reviewer asks if we expect these non-experts to configure things like bgp and acl? it is a good point bc e.g. elderly ppl generally don't configure bgp and acl. instead of these groups, we can say entry-level developers, end-users, and small business owners} A natural solution is to enable users to interact with networks through natural language. Natural language (NL), which is universally understood, does not demand any domain-specific expertise from the user. Although a handful of previous works have explored the use of NL for network management \cite{lumi, net2text}, \minlan{May cite lumi work here or home network configuration tools} recent progress in Large Language Models (LLMs) \cite{brown2020language, bommasani2021opportunities, wei2022emergent, chowdhery2023palm, li2025predictive} opens up greater possibilities to realize this vision. LLMs are already widely used to simplify other complex tasks, such as ChatGPT \cite{openai2023chatgpt} for drafting emails and GitHub Copilot \cite{github2023copilot} for coding, and the idea of general pretraining behind LLMs has also been applied to building foundation models for network dynamics \cite{dietmuller2022new, zhou2024netflowgen}.

In particular, LLMs have given rise to ``LLM agents,'' advanced AI systems that exhibit behaviors such as planning ahead, retaining context from prior interactions, utilizing tools, and adapting their behavior based on feedback \cite{llm_agents_guide, tian2025outlook}. By leveraging the reasoning capabilities of LLMs, agents have been used for a variety of complex tasks, including software development \cite{yang2024sweagent}, web browsing \cite{zhou2023webarena}, data science analytics \cite{luo2025assistedds}, and interfacing with operating systems \cite{wu2024oscopilot}. These advancements motivate the use of LLM agents for network control via NL \cite{wang2025intent}.

A few recent works have attempted to apply LLMs for network configuration \cite{vpp, nemoeval, netconfeval}. However, they largely focus on isolated tasks based on detailed inputs provided by expert network operators, which may not reflect real-world network scenarios. For example, NetConfEval \cite{netconfeval} proposes an LLM prototype that automatically retrieves relevant documentation for the specific networking use case, but the user is nevertheless expected to provide detailed instructions in the initial prompt, including a description of the topology and existing configurations on the network devices. We propose an agentic framework to reduce the expertise required from the user by augmenting an LLM agent with the domain-specific knowledge base and tools. \samuel{in contrast, our work is tailored for non experts, agentic framework to collect missing information from the user} \minlan{Cite existing LLM for network config papers, and say our focus is different} \samuel{for conext, we want emphasize that our focus is on building an agentic framework to interact with tools in a network simulation environment. we can also keep the point about accessibility}

% In addition to natural language as the medium for user-network interaction, 
With NL as the user-network interface,
we identify several important qualities of an accessible network management system: (1) \textit{The system should streamline configuration across different vendor languages.} Many networks are heterogeneous, consisting of devices from various vendors, each using its own configuration language. In general, pretrained LLMs are not trained on this domain-specific knowledge. (2) \textit{The system should be aware of real-time network conditions.} Networks are highly dynamic; devices join and leave, traffic patterns shift, and policies are frequently updated. While the knowledge base of an LLM is generally frozen after pretraining or fine-tuning, our system must continuously learn and track changes to the networking devices in real time. (3) \textit{The system should be able to incorporate external feedback.} For example, it is well-known that LLMs are prone to ``hallucination,'' \cite{xu2024hallucination} a phenomenon in which the LLM generates plausible but factually inaccurate information. This poses a significant challenge when applying LLMs to network configuration, which often has strict accuracy and reliability guarantees. To minimize disruptions from misconfigurations, the system might rely on feedback from users or external verifiers. (4) \textit{The system should facilitate user-friendly and efficient human-AI interaction.} It should be easy for users to verify configurations, further clarify their intentions, and interact with the AI system in an intuitive manner. This would enhance the overall user experience and make network management more accessible to a wider audience. \minlan{There are some redundancy of the following arguments and the arguments above. Please merge these two paragraphs into one}

This paper proposes a comprehensive agentic framework that addresses these challenges by (1) leveraging a vendor-agnostic intermediate representation (IR) such as YANG \cite{yang} to streamline configuration generation, (2) retrieving from a dynamic memory store, and (3) providing a unified interface for external feedback. We also (4) collect data from real users and develop a visualization tool to facilitate large-scale data collection. \joe{add (4) about data to correspond to the challenge (4) above to make the contributions very aligned with all identified challenges}  We conduct preliminary experiments that evaluate the ability of our system to accurately retrieve relevant details from the network state and to produce configurations that adhere to the IR specifications. Our experiments reveal that configurations are highly dependent on the accuracy of the retrieval results, verification is crucial to mitigate errors, and leveraging an agentic design improves the system performance. Based on these preliminary results, we call upon the research community to investigate retrieval techniques tailored to networking data, build robust verifiers for LLM integration, and standardize efforts to benchmark NL data in this domain under a shared IR. 

% In this paper, we propose a comprehensive system design that addresses these challenges by (1) implementing real-time dense retrieval of the network state, (2) leveraging a shared 
% YANG data model as 
% intermediate representation (IR) such as YANG \cite{yang} to express configurations across a variety of common networking tasks under the same semantics, (3) integrating mechanisms for automatic and human verification, and (4) collecting preliminary data from real users and developing a visualization tool to facilitate large-scale data collection. \joe{add (4) about data to correspond to the challenge (4) above to make the contributions very aligned with all identified challenges} We conduct preliminary experiments that evaluate the ability of our system to accurately retrieve relevant details from the network state and to produce configurations that adhere to the IR specifications. We observe that configurations are highly dependent on the accuracy of the retrieval results, verification is crucial to prevent errors, and decomposition of these components into multiple steps improves the LLM performance. Based on these preliminary results, we call upon the research community to investigate different methods for network state retrieval, build robust verifiers that verify not only the configuration syntax but also relevant network invariants, and standardize efforts to benchmark NL data in this domain under a shared intermediate representation. 

\joe{Outline:\\
- problem: network control/management/configuration (background, importance, challenges); focus on the difficulty for non-experts, so network operation is mainly done by experts\\
- our vision: more accessible control to non-experts, such parents in a home network, etc. and a natural way is to interact with natural language. (challenges: NL ambiguous, varibility) --> Progress in LLMs opens up possibilities to realize this vision, as these models are xxx like ChatGPT, co-pilot for coding.. so applying LLMs to enable accessible network control becomes appealing. \\
- (system design desireta) to enable a truly accessible system with which layman users can directly talk to the network to achieve what they want, we propose/envision different components of the systems that require better designs.. ideal system components with AI/LLM-integration driven: dyanmic network state incorporation (since LLMs are static and do not have network knowledge; for better contextualization of user request, verfication, more accurate configuration based on user request), automatic verification (non-expert might make mistakes, LLMs might make mistakes, and network configuration is less tolerant to mistakes), friendly and efficient human AI interaction (easier for human verification, further user intent specification, etc.).\\
- (more technique side) To achieve such systems, there are unique challenges in applying LLMs for network control. Challenges: LLMs are stateless, but networks are dynamically changing and configuration depends on network states (so we need states); network configuration langauge depends on application scenarios and hardwares, but LLMs may not have such knowledge of config-specific langauges (so we need a good IR to reduce the variability, and focus on NL-> IR development); LLMs make mistakes (or LLMs not good at precise semantics, etc. they hallucinate as an intrinsic drawback), or naively applying LLM can solve simple cases, but we have high accuracy requirements (so we need verification, and also human-ai interfaces for human verification); data side (hard to acquire data in the wild)\\
- (what we do) In this work, we propose a comprehensive system design xxx We conduct preliminary experiments to xxx Our key observations (e.g. LLMs are good but verification is veyr much needed; states are very important for dynamic network information, we have detailed analysis of data collected from real network operators, etc.), contributions, conclusions, calls for future works (still data and benchmarks, better retrieval, better verifiers, etc.).
}

\samuel{iterative prompting + visualization tool, LLMs are not good at complex tasks in unfamiliar network settings}

\section{Motivation and Challenges}
\label{sec-motivation}

\minlan{overall, all the bullets in this section do not go deeper than intro. If you compare sentences, they are almost equal}

\samuel{to illustrate network state challenge, be more specific, give estimates for the state sizes (use size of YANG Files, e.g.,), perhaps better not to limit ourselves to the example in the NL query, maybe cite sizes of config files * size of repr network}

\samuel{the Campion paper has a section that might be useful for formally explaining difference between vendor languages}

\samuel{so one way to differentiate our verification section from the Hotnets paper is that we incorporate retrieval in our verification. This also motivates agentic framework because the LLM can use the same retrieval tool early on to get network state, and again later for verification.}

We motivate our agentic design by highlighting several challenges of applying LLMs to network management. 

% \subsection{Managing Network State}
\smallskip

\bfsection{Handling Multiple Configuration Languages}
Large networks typically include equipment from various vendors. Managing these networks is complicated because different vendor languages implement configurations with vastly different structures, even for ostensibly similar functionalities \cite{campion}. To demonstrate, consider configuring MPLS administrative groups to include certain links (red) while excluding others (blue) in Juniper Junos and Cisco IOS. In Juniper, administrative groups are given explicit names (colors) \cite{juniper_mpls}:

\begin{verbatim}
mpls {
    admin-groups {
        red 15;
        blue 8;
    }
}
label-switched-path red-notblue {
    to 10.0.1.1;
    admin-group {
        include red;
        exclude blue;
    }
}
\end{verbatim}

In Cisco IOS, the same goal is achieved through a combination of affinity bits and masks \cite{cisco_mpls}:

\begin{verbatim}
interface Tunnel0
description red-notblue
ip unnumbered Loopback 1
tunnel destination 10.0.1.1
tunnel mode mpls traffic-eng
tunnel mpls traffic-eng affinity 0x08 mask 0x0A
tunnel mpls traffic-eng path-option 1 dynamic
\end{verbatim}

In addition to vastly different syntax, vendor-specific differences can lead to semantic drift in heterogeneous networks. In Juniper, LSP constraints use intuitive directives like \verb|include| and \verb|exclude|, allowing a logical OR operation for included colors. Cisco IOS, however, uses affinity bits and masks, which implements a stricter logical AND operation. The default behavior is also different between the two vendors. LLMs are generally not trained on these inconsistencies, and it is labor-intensive to fine-tune an LLM on all the nuanced differences among a large number of vendor languages. 

\minlan{Similar to 2.1, elaborate more on the actual challenges in more depths.}

\bfsection{Managing Network State}
The intention of a user query often depends on the real-time network state. For example, the LLM might need to know the current topology, the names of interfaces on each device, the IP addressing scheme, the VLAN assignments, the routing protocols in use, and any specific policies or constraints (e.g., access control lists, bandwidth limits, or high availability requirements). This requires a large amount of effort and technical expertise from the user, who must collect all the necessary contextual details to craft a comprehensive prompt. 

This challenge is further complicated by the fact that topologies \cite{ zhou2020automating} and network policies often change between queries, meaning that even minor updates to the network state could render previous prompts outdated. Moreover, this approach does not scale well. For example, we can estimate the complexity of configuring a SONiC switch \cite{sonic} by counting the number of leaf nodes in their YANG network models, where each leaf node corresponds to a configurable state variable in the switch. Then a simple switch that supports only interface configuration, ACLs, port channels, VXLAN, static routing, and route maps would have over 150 configurable parameters, and a modest network with 8 switches would already have over 1000 parameters to track \cite{sonicyangmgmt}. In a data center setting, state management may involve tens of thousands of network devices with many more state variables \cite{robotron, statesman}. 

In practice, prompt engineering can be challenging, and the accuracy of LLM outputs is highly dependent on the quality of the prompts provided. \samuel{joe can you give a source for this lol} Instead, we seek a scalable approach to state management that can adapt to changing network conditions and also reduce user effort.

% The intention of a user query often depends on the real-time state of the network. For example, consider a request to configure BGP:
%\begin{quoting}
%\texttt{Configure iBGP between Leaf and Spine devices.\\ Configure eBGP between each leaf and end-hosts.}
%\end{quoting}
%\vspace{0.1cm}
%Several ambiguities
\minlan{I don't understand what these ambiguities are} 
%arise from a lack of knowledge of the current network state. Here, the absence of information on existing BGP sessions and peering relationships can lead to misconfigurations. 
% While an LLM-driven system could require the user to provide these details, such an approach would place an excessive burden on a non-expert user. 
%
%The types of network states we need to consider in an LLM-driven system are extensive. For example, these may include topology information (nodes and links), node and link states (interface status, link utilization), traffic information (real-time traffic patterns and measurements), relationships (which node is associated with which link, VLAN assignments), device-specific configurations (routing tables, ACLs, firmware versions), and security settings (firewall rules, intrusion detection statuses).
%In a data center setting, state management may involve tens of thousands of network devices with exponentially more device-specific parameters \cite{robotron}.
\minlan{exponential to what? quote more detailed numbers of state sizes?} 
% Moreover, state variables are tightly coupled across devices, and accounting for these dependencies further complicates state management
\minlan{I don't understand, please elaborate more} %\cite{statesman}.

% \subsection{Ensuring Configuration Accuracy}

\bfsection{Ensuring Configuration Accuracy}
% It is well-established that LLMs tend to hallucinate,
Given that many networks require high accuracy guarantees, a robust network management system must mitigate LLM hallucination by implementing strong mechanisms for error detection and correction before deploying configurations. Network operators currently leverage a wide variety of specialized tools to reduce errors. These range from control plane analysis tools \cite{batfish} and semantic verifiers \cite{campion} to simulation tests. Our goal is to improve LLM reasoning on complex tasks \cite{feng2025unraveling} by enabling the LLM to act autonomously based on external feedback. To reduce the manual effort of integrating a large number of tools, we envision an LLM agent that dynamically invokes the appropriate feedback tool based on the present task. 

% Similarly, non-expert users may introduce inaccuracies due to a lack of technical knowledge. For example, they may underspecify their requests or ask for configurations not possible given the current network state. Consider the following request to configure ACLs:
% \begin{quote}
%     \texttt{Configure acls on each leaf node filtering unwanted traffic. Allow only certain hosts behind leaf0 to talk to hosts behind leaf1. This is done by specifying the allowed source and destination IP addresses within the acl.}
% \end{quote}
% \begin{quoting}
%    \texttt{Configure acls on each leaf node filtering \\unwanted traffic. Allow only certain hosts \\behind leaf0 to talk to hosts behind leaf1. \\This is done by specifying the allowed source \\and destination IP addresses within the acl.}
%\end{quoting}
%\vspace{0.1cm}
%Observe that the user did not specify which IP addresses correspond to the ``certain hosts'' behind Leaf 0 and Leaf 1. To generate an accurate configuration, the LLM must infer the missing information from the network state or solicit further clarification from the user. The LLM must also evaluate whether the requested configurations are even feasible given the current network state. If certain hosts behind Leaf 0 or Leaf 1 are not reachable, or if the specified traffic filtering rules conflict with existing policies, the system should provide feedback to the user and suggest possible alternatives. 

\samuel{examples of 1) user underspecification or 2) something that is not possible in the current network state. also say it is well-established that llms hallucinate. }

\minlan{This one do we need to cite the hotnets paper on LLM verification and clarify the difference?}

% \subsection{Collecting Natural Language Data for Networks}

\bfsection{Collecting Natural Language Data for Networks}
The lack of real-world data for benchmarking and evaluation significantly hinders the development of NL-based network configuration systems. Previous efforts to compile such a dataset have been insufficient. NetConfEval \cite{netconfeval} compiled a synthetic dataset based on template-based approaches, while other works have focused on task-specific scenarios \cite{nemoeval} or domain-specific contexts \cite{lumi}. These datasets all fall short of capturing the variability and complexity of real-world network configuration request. In order to develop and evaluate a system that enables accessible network control for non-experts in the wild, we need to establish benchmarks with NL utterances that reflect how users would actually express them in real-world scenarios.

Collecting such data is challenging because NL is not traditionally used for network management. Without existing systems to facilitate these interactions, it is difficult to predict how users would naturally express network configuration requests. Therefore, an initial system must be developed to observe and gather these interactions. In the following sections, we will discuss our methods for creating effective simulation and utilizing visualization tools to support large-scale data collection and system development.

\joe{lack of data for evaluation/benchmarking, also hinders the development of such NL-> config systems.\\
current systems attempting NL->network configs are simplistic, such as Lumi, CoNEXT paper: their NL utterances are synthetic (e.g. templated (NetConfEval), task-specific (NeMoEval), domain-specific (Lumi)).\\
but to enable a capable system in the wild for widely accessible network control to non-experts, we need to establish benchmarks (at least) with natural langauge network control requests that resemble actual human utterances. (motivating from the necessity of real NL data)\\
challenges in collecting such data: there has not been any work/system for this, and tradionally ppl do not use NL for network configuration, we need to shift the paradigm with new data and system; collecting such data is also hard, as we don't know how people would talk to the networks without building such a system, and for good data collection, we want to mimic the use cases as real as possible. (so we can talk about our data collection setup, questionnaire guidelines, example, later)\\
--> visualization tool, also good for larger scale real data collection for system development. (can touch briefly, but detailed discussions come in later sections)}

\presec\section{Proposed System Components}
\label{sec:sys-components}

\begin{figure}[t]
    \centering
    % \prefig
    \includegraphics[width=0.45\textwidth]{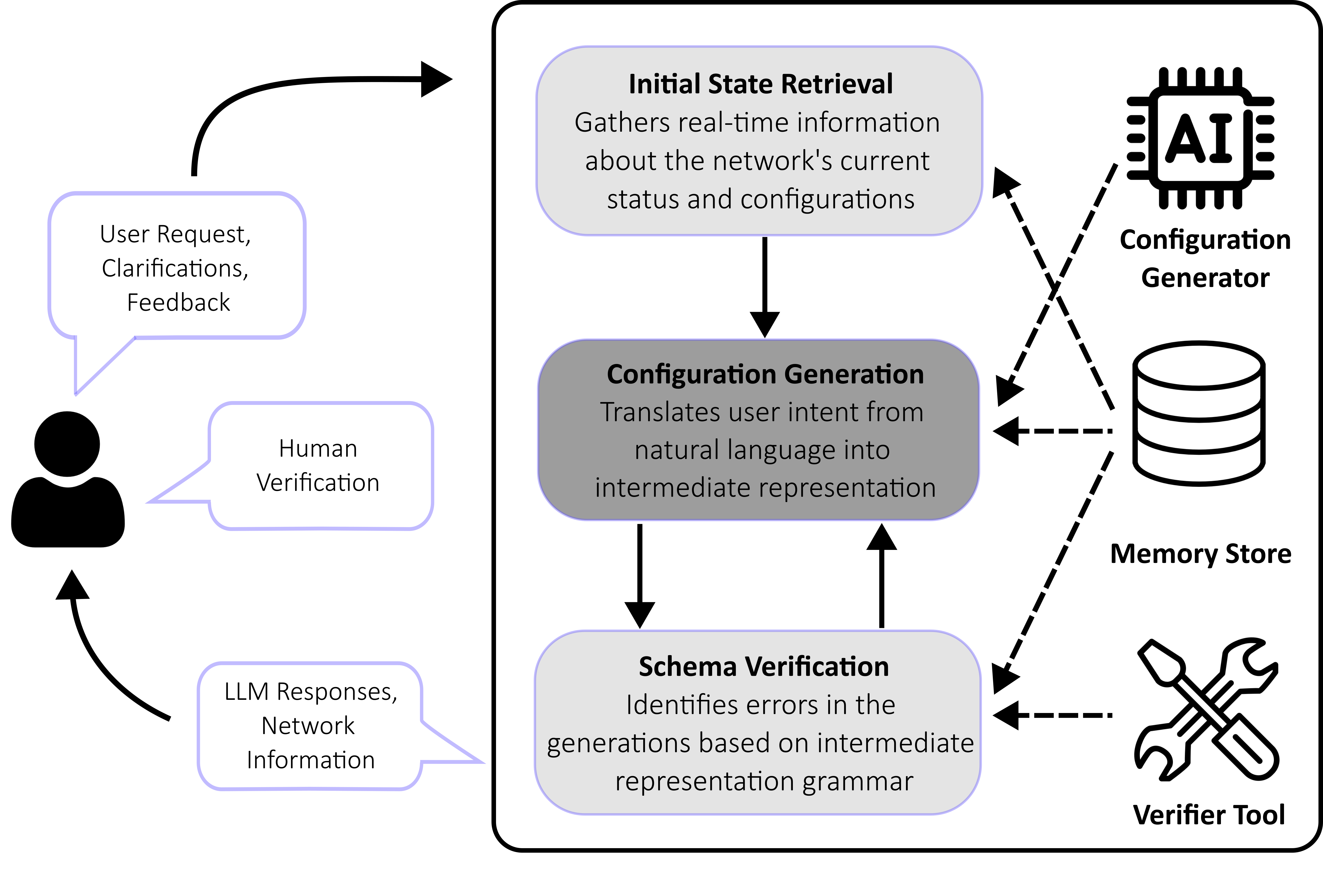}
    \prefigcaption\caption{System overview for LLM-based automatic IR generation from natural language.}
    \postfigcaption
    \label{fig:sys-components}
\end{figure}

To tackle the challenges discussed in \S \ref{sec-motivation}, we propose the design of a systematic, agentic framework. The framework consists of a core decision-making LLM, which we refer to as the \textit{LLM agent}. This agent is augmented by a configuration generator (\S~\ref{subsec:IR}), a memory base (\S~\ref{subsec:state-retrieval}), and an interface for external feedback (\S~\ref{subsec:feedback}). Figure \ref{fig:sys-components} provides an overview of the system components, as well as how human users interact with them. An example end-to-end workflow from our agentic system is provided in Appendix \ref{sec-appendix-a}.

% We also introduce a visualization tool to facilitate human-agent interaction. 

\samuel{systematic, agentic framework design. we have one realization of this system. for the retrieval, we have a separate module that is part of the agent, so it is not just a single agent. we propose this framework and here is one realization of it. we can have specialized models or more advanced components. }

\samuel{interleave external feedback in the reasoning process. }

\samuel{think about agentic framework. decision making module, at the core is the llm that does the translation. This is the core part of the agent. There is also a memory component that stores the state and the defn of the IR. These are two databases. And this is connected to decision making through retrieval kind of process. and third component is feedback/verification. }

\samuel{call it core decision making agent, memory agent (retrieval), verification agent, and this is connected to generation module through this feedback. We use compiler to check errors but other things can check errors, like even another LLM. Finally we have this human interaction component outside of this. This is the interface from human as entry point to this system. }

\samuel{from the surface they seem very disjoint. we are talking about different pieces without an overall map to tie them togeether. so we formerly attached the components to each of the challenges. }

\samuel{instead of learning different agents for different languages, we want to a single unified agent for all these things, this lowers the barrier for developing new systems. so we use aN IR to facilitate this. this is one way to formulate it. break down problem to fit it in scale. having an agentic system that can tackle this automatic confgiuration from NL we want it to be scalable reliable dynamic, memory retrieval, we also need user interface so this is transparency. }

\samuel{in the section we need smtg to connect ot later section on agentic framework. mention smtg like memory. this calls for the need of a memory component. or need verification for reliability. we can also rearrange order. building intelligent system is harder for different langauages bc we need to train different languages models or different parsers. we can focus on one IR to focus on making this aganet good and using tools to translate from IR to trrget language. connect this to what we do later. and order can be changed. }

\samuel{fir the layman user to check, we have this interface so user doesnt have to check technical details, they can just check if their goal was achieved, and visualization tool is to facilitate this. lower barrier on user side.}

% We now turn our attention to the development of a stateful, error-free, and user-friendly LLM system that translates natural language requests into the IR. Figure~\ref{fig:sys-components} shows the various components of the LLM system, as well as how human users interact with it.

% To tackle the challenges discussed in \S \ref{sec-motivation}, we describe an LLM integrated system aimed at being ideally stateful, versatile, error-free, and user-friendly.
% Figure~\ref{fig:sys-components} provides an overview of core system components, as well as how human users interact with them.

\presub\subsection{IR for Configuration Generation}\postsub
\label{subsec:IR}

The first component of our system is a \textit{configuration generator} that translates an NL request into configuration code using a separate LLM. As discussed in \S \ref{sec-motivation}, it is challenging to build an intelligent LLM system that can operate reliably with multiple vendor languages. Each new vendor language would require developing a unique parser to ensure interoperability with other system components, as well as fine-tuning a separate vendor-specific LLM model. 

\begin{figure}[t]
    \centering
    \includegraphics[width=0.45\textwidth]{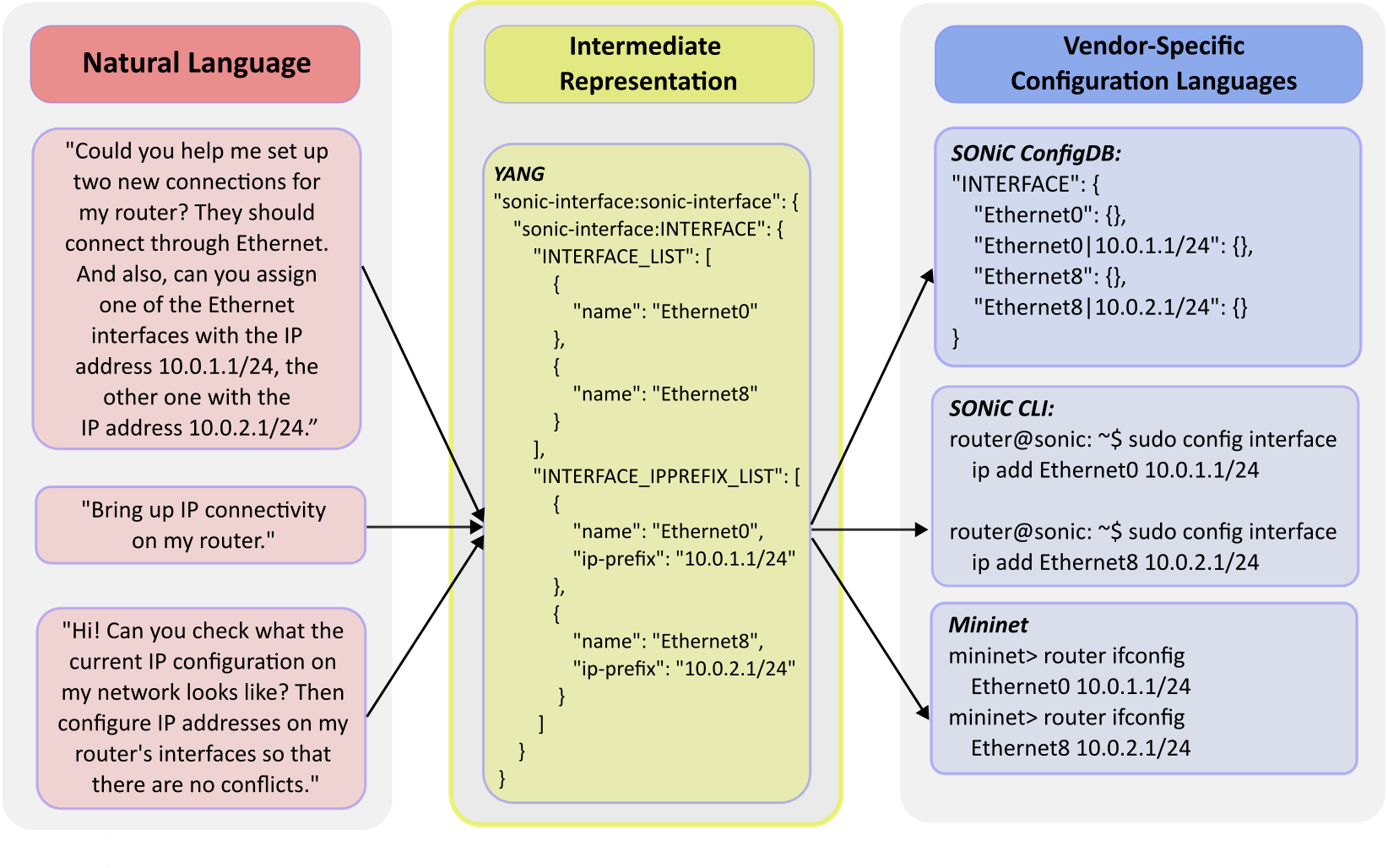}
    \prefigcaption\caption{Using a unified IR to bridge diverse natural language requests and application-specific network configuration languages.}
    \postfigcaption
    \label{fig:NL-IR-config}
\end{figure}

We introduce the use of an \textit{intermediate representation} (IR) to generate configurations. The IR is a vendor-agnostic abstraction that distills the complexities of NL requests into a structured format that can then be deterministically compiled into low-level network configuration code. Figure \ref{fig:NL-IR-config} illustrates our vision of a single IR through an example configuration. \minlan{We need a comprehensive description of the IR instead of an example, because this is a design section. Also, what's the key innovation/insights in your IR design. What's a naive solution and why your solution is better?} In this pipeline, we first map user requests from NL to the IR. NL expressions are highly variable, and a single configuration objective can be expressed in any number of ways. Although precise interpretation using rule-based methods is difficult, LLMs are capable of handling NL variations. Next, we map the configuration from the IR to various vendor-specific languages. Existing synthesizers already achieve this via explicit one-to-one mappings between elements in the IR and their counterparts in the target language.

Rather than build multiple LLM agents that are individually tailored to specific vendor languages, our approach develops a single intelligent LLM for configuration generation that is fluent in the IR and augments it with compilers to convert between the IR and the target language. Since only one LLM is used, it is easier to improve the configuration generation by experimenting with different prompts or by fine-tuning a single LLM, as done in \cite{netconfeval}. In designing the IR, we seek a \textit{vendor-agnostic}, \textit{structured}, and \textit{expressive} abstraction. A structured representation, as opposed to a CLI-based language, facilitates schema validation, which can identify missing fields, incorrect data types, or out-of-range values in the LLM output. Moreover, an expressive IR helps to support a wide range of network configurations, including those that are proprietary to specific vendors. 

We use YANG \cite{yang} as our IR in this work. Figure \ref{config-generator-prompt} shows the prompts that we used for configuration generation in YANG. YANG follows a structured hierarchical schema that can be extended to a wide range of network tasks, and a growing number of vendors have added YANG support on their devices.

% \textit{Previous version}

% We introduce the use of an \textit{intermediate representation} (IR) as a fundamental part of our approach to network configuration via NL.
% As discussed in \S \ref{sec-motivation}, network operators are often required to be familiar with many configuration languages from different network vendors. However, creating separate LLM systems for each language would complicate the overall user experience, since users would have to interact with multiple specialized systems.
% As discussed in \S \ref{sec-motivation}, network operations are often conducted in vendor-specific configuration languages, complicating automated pipelines for multi-vendor LLM systems.
% Instead, we use an IR to distill the complexities of NL requests into a structured, vendor-agnostic format that can then be deterministically compiled into low-level network configuration code. 

% Moreover, by establishing a deterministic mapping between a unifying IR and various vendor-specific languages, we can focus our LLM system to only translate natural language expressions into the IR.

\presub\subsection{Memory Base}\postsub
\label{subsec:state-retrieval}

We augment the LLM components with a \textit{memory base}, implemented as two separate data stores: one for the network state and another for the IR documentation. The \textit{network state} is dynamically refreshed at query time to provide an up-to-date understanding of the network. The \textit{IR documentation} provides the configuration generator with the IR schema.

In a memoryless approach, the user writes all the necessary context directly in the prompt via a template. It is left to the user's judgment to determine the relevant context for their specific use case. This alternative works well for one-off queries where the required knowledge can fit within the prompt’s token limit, and iterative interactions are limited. However, for network management, an agentic design with a decoupled memory component is preferable due to its greater dynamism and scalability \cite{li2025context}. The memory component can also be flexibly expanded to include additional resources. For example, future work could include a repository of previous interactions between the user and the LLM agent, which would improve its performance in iterative workflows.

The agent leverages dense retrieval on a vector database \cite{karpukhin2020dense} to fetch the most relevant information from the memory base. The configuration and topology files, as well as the IR documentation, are first encoded into vector representations via a pretrained embedding model to capture the semantic information of the data. Similarly, user queries in NL are also embedded into vectors, and retrieval is done by finding the closest matching vectors in the database. Vector databases scale well and enable fast querying, making them suitable for low-latency, large-scale network environments. Future work could improve the retrieval mechanism based on the hierarchical nature of networking data.

% \bfsection{Two-Stage State Retrieval}
% We use separate LLM agents in a two-stage retrieval process of the network state. Given a network request, the first LLM agent performs a feasibility check to ensure the request can be executed given the current network state. If discrepancies or missing details are found, this agent provides feedback to the user and prompts for additional information or adjustments to the request. Once the configuration is deemed feasible, the second LLM agent identifies and retrieves parts of the network state that are relevant to the user request. By default, the LLM agent\minlan{Do you mean the first agent or the second one? give each agent a name}outputs the retrieved parts directly in the original format of the configuration files. However, we find that future configurations are better reflecting the network state when we instruct the LLM agent to describe these retrieved parts in NL instead.\minlan{How is this sentence relevant to the challenge in sec 2 bullet 1?}  \joe{clarify the sentence a bit}
\minlan{Overall, why two agents are the right approach? what are the alternatives? What are the tradeoffs?}
% \smallskip\newline
% \vspace{-0.01in}

% \bfsection{Dense Retrieval Mechanism}
% Both LLM agents leverage dense retrieval on a vector database \cite{karpukhin2020dense}. 
% Existing network configuration and topology files are first encoded into vector representations via a pretrained embedding model, capturing semantic information of the data.
% Similarly, user queries in NL are embedded into vectors too, and retrieval is done by finding the closest matching vectors in the database.
% Vector databases scale well and enable fast querying, making them suitable for low-latency, large-scale network environments. Further optimizations could be made by performing incremental updates to maintain a consistent database and leveraging the hierarchical nature of networking data. 
\minlan{We will work on further improve the retrieval based on the hierarchical nature of networking data}

\presub\subsection{Feedback Integration}\postsub
\label{subsec:feedback}

We design an interface to integrate \textit{external feedback} into our system. Our design is similar to \cite{yao2022react}, in which the LLM agent interleaves reasoning action generation. An agentic feedback loop is more flexible than a predefined feedback workflow because it allows the LLM agent to react autonomously to feedback. For example, Figure \ref{feedback-prompt} shows the prompt for an agent with access to a verifier. Based on the verifier's feedback, the agent may leverage other tools or retrieve more details from memory before regenerating the configuration. 

The user is another valuable source of external feedback. We explicitly incorporate entry points for human feedback at critical steps in the workflow. During the initial retrieval of the network state, the agent iteratively clarifies with the user that their intent is clearly understood and aligns with the retrieved information from the network state. Before invoking the configuration generator, the agent may ask the user for missing details. Finally, after automatic verification, the agent presents the configuration to the user for review. This effectively results in a continuous user-agent dialogue, which improves the overall accuracy of the system and eliminates the need for an exhaustive prompt from the outset.

% \textit{previous version}

% To build a reliable LLM system, we employ a framework for modular verification \cite{lightyear, vpp}, which decomposes the complex task of ensuring overall correctness into smaller, more manageable components. 
% We identify the following criteria for correctness, although this list may not be exhaustive: (1) the user request is feasible within the network state, (2) the output configuration reflects the network state, (3) the output configuration follows the IR's syntax specifications, and (4) the output configuration aligns with the human's intention. We address (1) and (2) by implementing the state LLM agent in \S \ref{sec:state-retrieval}.
\minlan{Since 1 and 2 are already discussed, maybe just focus on 3  in this bullet and elaborate more; also discuss 4 in the next bullet. Remove the modular verification argument} 

\joe{we can make this more stand out, by going into the visualization quickly or having a separation here}

\begin{figure}[t]
    \centering
    \includegraphics[width=0.45\textwidth]{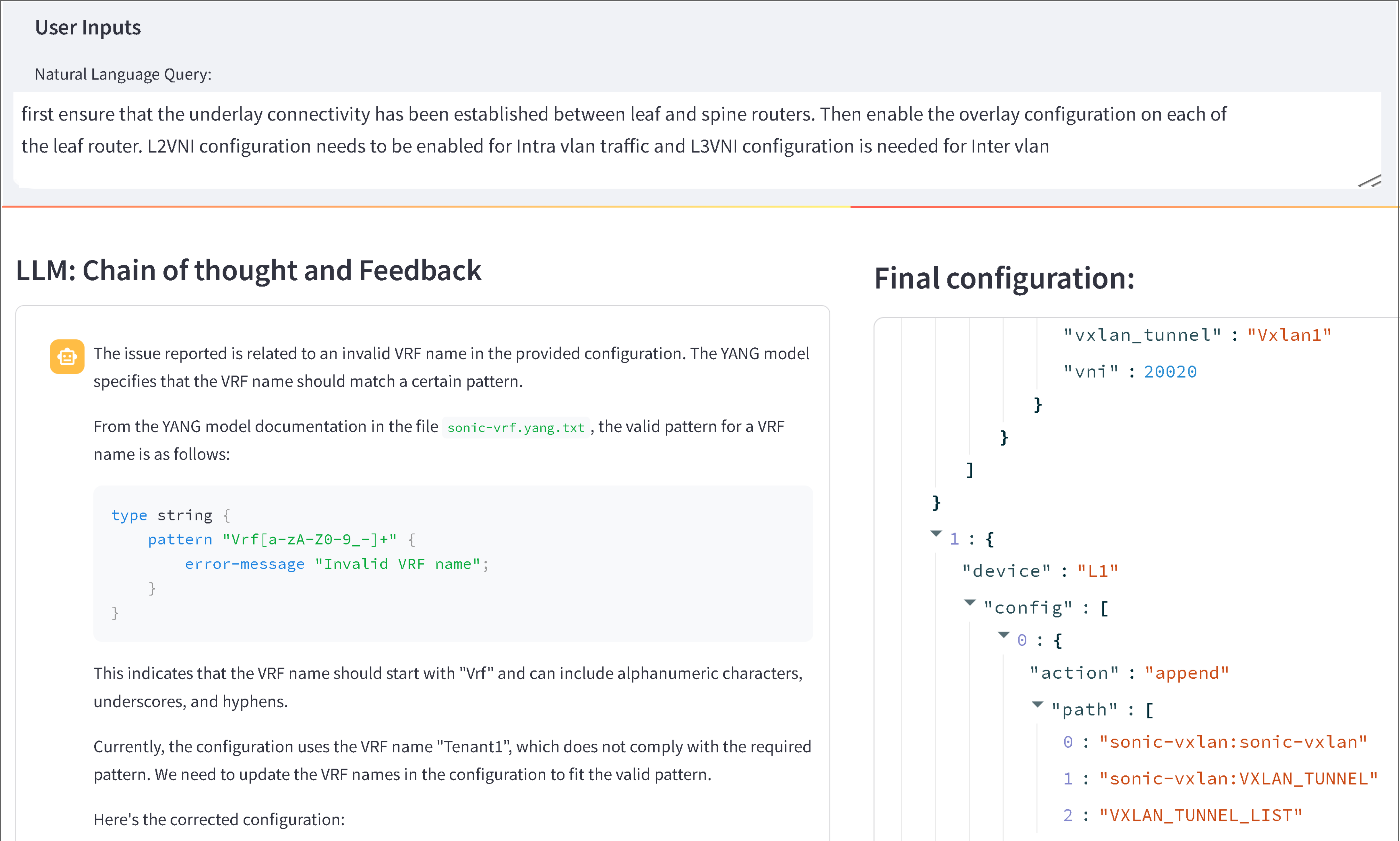}
    \prefigcaption\caption{Visualization interface for user-system interaction
    }\postfigcaption
    \label{fig:visualization}
\end{figure}

% \section{Data Collection and Visualization}
\presec\section{Data and User Interface}\postsec
\label{sec:data-visual}

\minlan{I think we can remove this section and just say one paragraph of data collection for experiments, if we are short of pages}
\joe{I think one idea is to merge this into experiment sections 2, to describe data collection and visualization as a step for the preliminary experimental study towards a real system}

\bfsection{Data Collection} Real-world data is critical to the development and evaluation of NL-based network control systems. We are the first to collect NL utterances that captures the distribution of configuration intents as they occur in real-world scenarios. %These datasets are mostly synthetic and template-based \cite{netconfeval}, and fail to capture the variability and nuances of real-world user requests. \joe{we can again touch upon previous works on their NL data, why it is necessary to collect data like ours} 
We collected a small dataset by distributing a survey to network operators from a large networking solution provider.\joe{should we say Cisco? probably not. maybe ask Minlan}\minlan{maybe just say network operators from a large networking solution provider in the  paper} Our survey presented a 3-stage Clos network scenario \cite{ciscolearninglab} and solicited network configurations in NL for several common management tasks. %: IP, BGP, VXLAN, Port Channels, OSPF, QoS policies, and ACL configurations. 
For each task, we asked the respondent to describe a network-wide policy in NL and provide per-device configuration requests in NL. %, and rate their familiarity with that task on a scale of 1 to 5. 
The per-device configuration requests are used as the inputs for our LLM, while the network-wide policy descriptions are used to evaluate the LLM's outputs against the user's intentions.
%We gauge the confidence level in the responses using the familiarity ratings. 

% In our dataset, the average familiarity rating across 13 collected NL utterances was 4.153 out of 5, suggesting that participants had a strong understanding of the network management tasks. 

\samuel{Thoughts:\\
data challenge: we want data in the wild, i.e. reflects distribution of real world natural language utterances, yadayada. 
we should probably also talk about the data that we DID collect. 
What do we do next? We build a visualization tool, we want VISUALS, maybe mention home network topology, you can see nl input, intermediate explanations, network connections, etc. etc., which we believe will help facilitate large-scale data collection. We want to get the types of user utterance that we would get if they had the desired system. Like Chatgpt, we want to deploy something and improve it while at the same time collect (better) user data. (thumbs up thumbs down in the interface itself). 
and show a screenshot of our visualisation tool. maybe at least two paragraphs here, some sentences or overview about data collection, and then transition to what we do (3 clos network) and also append our guidelines for the user, and then transition to the visualization tool.} 

\joe{echoing the data challenge. better to collect data from the wild, to have real user utterance should they be given the desired system.\\
--> talk about what we do for the challenges \\
visualization tool. can help interact with the system, a simulation environment (best if also be real, chatGPT example). and visual tool can also helps data collection in the wild.
--> talk about what we built as a preliminary network config visualization tool with NL. can have a screenshot of our system.}

\bfsection{Visualization Interface} To facilitate data collection and system evaluation, we design a conversational interface with which non-experts can view responses from the agent and give further instructions. Our aspiration is akin to that of ChatGPT's success story, in which the system is continuously deployed to a wide audience and refined based on user interactions. In the same spirit, our visualization tool features an intuitive interface to facilitate user-agent dialogue.

% % To simulate an AI system in the wild, our survey instructed respondents to envision practical, real-world scenarios, and to communicate their intent as naturally as possible, as if they were speaking to another person (as opposed to a bot). Nevertheless, we recognize that a survey, being a controlled setting,  does not faithfully reflect how people would actually interact with an LLM system.
% Despite our best effort to collect real NL utterances by instructing users to express their intents as naturally as possible as if they were speaking to another person (as opposed to a bot), a more realistic application environment is still desired.
% To that end, we design a visualization tool to simulate this user experience and to collect this data at scale. Our aspiration is akin to that of ChatGPT's success story, in which the system is continuously deployed to a wide audience and refined based on user interactions.
% It simulates the user experience with our desired AI system in the wild with intuitive user interfaces.

An example interaction is shown in Figure~\ref{fig:visualization}, where the user wants to set up an overlay network to extend VLANs and enable intra- and inter-VLAN traffic. All intermediary responses from the LLM agent are displayed in NL. This allows the user to monitor the agent's reasoning and engage with it in continous dialogue. Future work can improve the interface by providing additional network details, such as a visualization of the network topology. With this visualization tool, we hope to make the network management system accessible to a large audience, and in doing so, enable large-scale data collection for future development.
\joe{cite the figure somewhere}

% We integrate multi-round prompting at key stages to engage the user in a continuous dialogue with the network management system. Not only does this improve the overall accuracy of the system, but it also reduces the need for exhaustive details from the outset, making it more accessible for non-experts.  During the initial state retrieval, the system engages in iterative clarification with the user to confirm that their intent is clearly understood and accurately reflected in the network state. In the configuration generation phase, the LLM may ask for further details if it lacks sufficient information. Finally, after automatic verification, the system presents the configuration to the user for review and feedback. To facilitate this dialogue between the user and the system, we design an intuitive conversational interface with which non-experts can readily view responses from the LLM agents and give further instructions.
\joe{I'll add some references for human-AI collaboration}

\joe{echoing challenges and system visions we proposed earlier:\\
- how states are managed and utilized\\
- how we correct mistakes through (different levels of) verification\\
- iterative prompting/process, conversational interface; for easier feedback. LLMs can provide explanations in NL when things go wrong, user could provide feedback to LLMs in NL. Can cite George's verfication paper on humanizaer to NL and modularized verification.}

\presec\section{Experiments and Results}\postsec
\label{sec:experiments}

In this section, we present the results of our experiments. We test our LLM agent on two datasets: a pre-existing dataset, and the real-world network scenario dataset, as described in \S \ref{sec:data-visual}. For the first dataset, we compare our agent with a baseline system to underscore the advantages of configuration generation via LLMs. Then we run our LLM agent on our own dataset to demonstrate the effectiveness of its components in real-world network configuration scenarios.

% since the baseline system is unsuitable for the more complex requests in the second dataset, we run only our LLM agent and demonstrate the effectiveness of the agentic components. 
% In this section, we present results from our preliminary experiments. First, we we compare the performance of our system against Lumi \cite{lumi},\joe{a bit of context of what Lumi is; we could say "compare to a baseline xxx"} focusing on intent recognition, and show that our general-purpose LLM-based system performs comparably to an LSTM that was specifically trained on Lumi's target datasets. Next, we demonstrate the effectiveness of state retrieval and automatic syntax verification on the 3-stage Clos network dataset, as described in \S \ref{sec:data-visual}.

\presub\subsection{Comparative Analysis with Lumi}\postsub

We compare our LLM agent with Lumi \cite{lumi} as the baseline. Lumi is an intent-based network management system that translates NL intents into low-level configuration commands. It trains an LSTM to perform Named Entity Recognition (NER) on operator utterances. Using a system of templates and predefined rules, the extracted entities are then structured into configuration statements expressed in their Network Intent Language (Nile). 
We test on the \textit{alpha} and \textit{campi} datasets provided with Lumi. \textit{alpha} consists of 150 synthetically-generated intents, while \textit{campi} consists of 50 real-world intents obtained by crawling university websites. All intents involve network tasks common to campus networks, and each is manually annotated with ground-truth entities and Nile intent.
% Each natural language utterance is manually annotated with a list of ground-truth entities as well as the corresponding Nile intent.  

For evaluation, we adopt Lumi's metric on entity tagging with Precision, Recall, and F1-score measures.
% In Lumi, the authors evaluate their LSTM system on entity tagging by calculating Precision, Recall, and F1-score on the extracted entities. In replicating their evaluation, we consider an entity to be correctly extracted only if it perfectly matches the corresponding entity in the ground truth dataset, as labeled by Lumi's authors. 
Moreover, we would like to evaluate the accuracy of the generated Nile configurations. We propose the metrics ``exact match'' and ``fuzzy match,'' where the latter is more lenient, accounting for slight variations of text strings such as capitalization and synonyms of a consistent term. 
% An exact-match score returns 1 if the generated Nile statement perfectly aligns with the ground truth annotation. For fuzzy matching, we use a more lenient measure that accounts for slight variations. We preprocess each generation by normalizing text to lowercase and replacing synonyms with a consistent term. The score is then determined by the ratio of matching Nile operations and extracted entities to the total number of operations and entities in the ground truth statement.
This offers a more fine-grained evaluation of the generated Nile configurations. 

We instruct GPT-4o to convert the NL utterances into Nile intents based on Nile's BNF grammar. Since Lumi's LSTM system does not consider the network state, we make a fair comparison by only including the following components of system: configuration generator with one-shot prompting, and schema verifier with a BNF grammar verifier for Nile.  

\begin{figure}[t]
    \centering
    \prefigcaption\includegraphics[width=0.48\textwidth]{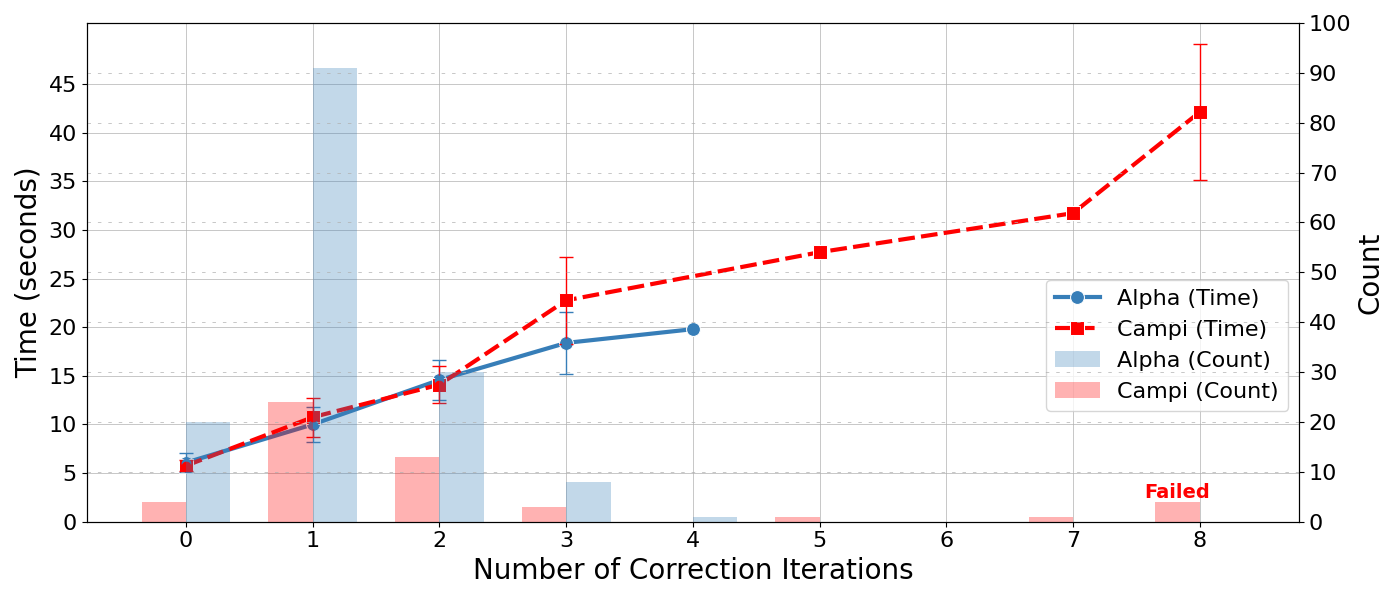}
    \caption{Analysis of Nile intent correction.}
    % \postfigcaption
    \label{fig:lumi-latency}
    \vspace{-0.1in}
\end{figure}

The results of our evaluation are given in Table~\ref{tab:lumi-table}. They demonstrate that our general-purpose LLM agent performs comparably to an LSTM that was specifically trained on these datasets. We believe that incorporating more sophisticated prompting techniques, such as multi-shot prompting, would further enhance the accuracy of our system. Finally, Figure~\ref{fig:lumi-latency} reports the amount of time taken by our LLM system with respect to the number of verification iterations. As shown in the figure, the BNF verifier is able to correct all syntax errors within 8 iterations for 195 out of 200 total intents. 

\begin{table}[t]
\centering
\resizebox{\columnwidth}{!}{
\begin{tabular}{@{} l cccc @{}}
    \toprule
    & \multicolumn{4}{c}{Entity Recognition Metrics} \\
    \cmidrule(lr){2-5}
    & \multicolumn{2}{c}{\textit{alpha}} & \multicolumn{2}{c}{\textit{campi}} \\
    \cmidrule(lr){2-3} \cmidrule(l){4-5}
    & LLM & Lumi & LLM & Lumi \\
    \midrule
    Precision & \textbf{0.84} & 0.79 & \textbf{0.65} & 0.55 \\
    Recall & 0.82 & \textbf{0.86} & 0.54 & \textbf{0.69} \\
    F1 Score & \textbf{0.83} & 0.82 & 0.59 & \textbf{0.61} \\
    \bottomrule
\end{tabular}
\quad
\begin{tabular}{@{} l ccccccc @{}}
    \toprule
    & \multicolumn{2}{c}{Match (\%)} \\
    \cmidrule(lr){2-3}
    & \multicolumn{2}{c}{\textit{campi}} \\
    \cmidrule(lr){2-3}
    & LLM & Lumi \\
    \midrule
    Exact & 8.00 & \textbf{22.00} \\
    Fuzzy & \textbf{75.83} & 73.38  \\
    \bottomrule
\end{tabular}
}
\smallskip
\caption{Comparison of system performances based on entity recognition (left) and Nile intent match (right).}
\label{tab:lumi-table}
\vspace{-0.1in}
\end{table}

\medskip

%\begin{table}[t]
%    \centering
%    % \resizebox{\columnwidth}{!}{%
%    \begin{tabular}{@{} l cccc @{}}
%        \toprule
%        & \multicolumn{2}{c}{\textit{alpha}} & %\multicolumn{2}{c}{\textit{campi}} \\
%        \cmidrule(lr){2-3} \cmidrule(l){4-5}
%        & LLM & Lumi & LLM & Lumi \\
%        \midrule
%        Precision & \textbf{0.836} & 0.786 & %\textbf{0.654} & 0.551 \\
%        Recall & 0.815 & \textbf{0.864} & 0.543 & %\textbf{0.694} \\
%        F1 Score & \textbf{0.825} & 0.823 & 0.594 & %\textbf{0.614} \\
%        \bottomrule
%    \end{tabular}
    
%    \bigskip
    
%    \begin{tabular}{lcc}
%        \toprule
%        & LLM & Lumi \\
%        \midrule
%        Exact Match (\%) & 8.00 & \textbf{22.00} \\
%        Fuzzy Match (\%) & \textbf{75.83} & 73.38 \\
%        \bottomrule
%    \end{tabular}
%    % }
%    \caption{Caption}
%    \label{tab:lumi-comparison}
% \end{table}

% These results demonstrate that our general-purpose LLM-based system matches the accuracy of an LSTM that was trained on these datasets. We believe that incorporating more sophisticated prompting techniques, such as multi-shot prompting, would further enhance the performance of our system. Finally, Figure 4 reports the amount of time taken by the LLM components with respect to the number of verification iterations. We note that the automatic verifier is able to resolve all Nile syntax errors within 8 iterations for 195 out of 200 total intents. 

% \medskip

\presub\subsection{Evaluation on 3-Stage Clos Network}\postsub

Next, we assess the agent's ability to generate complete and correct configurations for the 3-stage Clos network dataset. In our experiments, we use YANG as the IR and the Python library pyang as an automatic syntax verifier. We use \cite{sonicyangmgmt} to compile the YANG generations into SONiC. We populate the memory base with topology information and per-device configuration files. The backbone is deployed with GPT-4o, while we leverage OpenAI's text-embedding-ada-002 \cite{text-embedding-ada} model for the retrieval component. We limit the scope of our evaluation to all components that do not require human-in-the-loop and evaluate the correctness of the final configurations based on whether they are syntactically well-formed and accurately reflect user intent. Each generation is evaluated as either correct or incorrect based on human expertise. 
\joe{evaluation metric (human expert); run 3 variations, average acc}

\begin{figure}[t]
    \centering
    \includegraphics[width=0.48\textwidth]{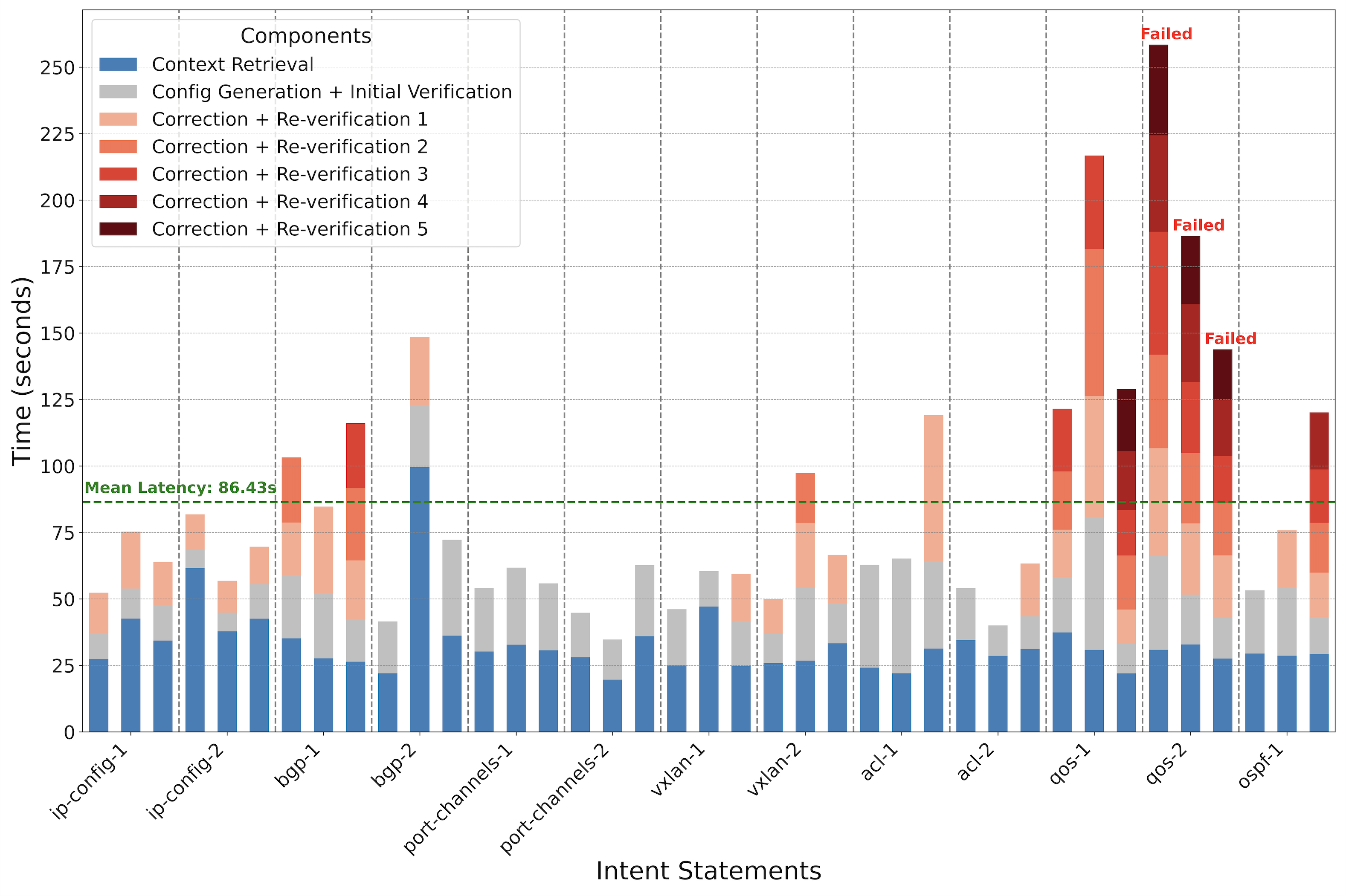}
    \caption{Latency analysis of system components including re-verification.}
    \label{fig:latency-detail}
    \vspace{-0.1in}
\end{figure}

We run each of 13 NL requests through 3 trials, resulting in 39 total generations. Table~\ref{tab:clos-dataset-table} presents our evaluation results. Without verification, only 33.33\% of the outputs are generated with proper YANG syntax, but our verifier corrects 87.5\% of these errors within 5 iterations. Among the verified outputs, 76.92\% can be deployed without errors. The remaining errors generally involve inconsistent IP and AS assignments, missing configurations for specific devices, or missing deny rules in ACL generations. These errors may be further reduced by incorporating feedback sources for verification. We also conduct an ablation study to demonstrate the effectiveness of state retrieval. Without the retrieval mechanism, the LLM exhibits significant hallucinations on every input. Including a retrieval mechanism that returns the relevant state components in the raw configuration file format results in only 46.15\% accuracy. However, instructing the LLM agent to describe the retrieved parts in NL rather than in the original configuration file format drastically improves accuracy to 76.92\%. Finally, Figure~\ref{fig:latency-detail} presents a detailed latency breakdown of the various components in the LLM system across all generations. We observe that on average, each query requires 86.43 seconds to complete. Moreover, more than 61\% of the trials require at least one round of correction and re-verification, underscoring the significance of feedback integration in our agentic system. 

\begin{table}[t]
    \centering
    \begin{tabular}{l c}
    \toprule
    Configuration Method & Accuracy (\%) \\
    \midrule
    Verifier Only & 0.00 \\
    State Retrieval (NL Output) Only & 33.33 \\
    State Retrieval (Raw Output) + Verifier & 46.15 \\
    \textbf{State Retrieval (NL Output) + Verifier} & \textbf{76.92} \\
    \bottomrule
    \end{tabular}
    \medskip
    \caption{Network configuration accuracy across different setups of the LLM-based system.}\posttablecaption
    \label{tab:clos-dataset-table}
    \vspace{-0.1in}
\end{table}

\joe{two experimental studies: one on Lumi data and comparing with Lumi, and the other one on real Cisco data. They have different data and evaluation.}

\joe{overview paragraph of experiments. and then subsections of two experimental studies. within each one, we talk about setups of data, eval metrics, and baseline/model. and then results, and then analysis.}

\samuel{show sample lumi and cisco utterance, different domains, but we test on the same LLM system and the concept works for both - IR is different but conceptually it is the same and it works}

\samuel{ablation study to justify state retrieval - with state and without state retrieval and compare results}

\section{Conclusion}
\label{sec-conclusion}

We present a vision to make network management widely accessible by enabling non-experts to interact with network systems through natural language. To realize this vision, we designed an agentic framework to facilitate state-aware, vendor-agnostic, interactive, and user-friendly network control. We also address the challenge of using LLMs to perform configurations in multiple target languages by introducing an intermediate representation. We hope that future work will focus on enhancing these core components of LLM-based network control systems, particularly by building verifiers for additional network invariants and developing more user-friendly interfaces to scale out data collection and realiable benchmarking.

% \vspace{1.25cm}

\newpage
\bibliographystyle{ACM-Reference-Format}
\bibliography{reference}

\appendix
\clearpage

% \section{End-to-End Example of IP Configuration}
\section{Example End-to-End Workflow for IP Configuration}
\label{sec-appendix-a}

We provide an example workflow where the user asks our agent to perform simple IP configurations on network devices. While the exact workflow varies between queries, the following illustrates the key interactions between the user, the LLM agent, and the system components. The query below is provided verbatim from our 3-stage Clos network dataset and exemplifies the type of NL utterances that we expect from real-world network operators during deployment. Responses are truncated for brevity.\\

\noindent \textbf{User}: \textit{Provides initial query.}
\begin{verbatim}
  Connect Ethernet4 of each Leaf to Ethernet4
  of each Spine and assign IP addresses to 
  them to bring up connectivity
\end{verbatim}

\noindent \textbf{Agent}: \textit{Retrieves network state from the memory store.}

\noindent \textbf{Memory Store}: 
\begin{verbatim}
  Spine0 (S0)
    Ethernet4: Interface up with speed 100000 
    Mbps and MTU 9100 [/S0/config_db.json]
  Spine1 (S1)...
  Leaf0 (L0)
    Ethernet4: Interface up with speed 100000 
    Mbps and MTU 9100 [/L0/config_db.json]
  Leaf1 (L1)...
\end{verbatim}

\noindent \textbf{Agent}: \textit{Retrieves IR documentation from the memory store.}

\noindent \textbf{Memory Store}:
\begin{verbatim}
  list INTERFACE_IPPREFIX_LIST {
    key "name ip-prefix";
    leaf name {
      type leafref {
        path /port:sonic-port/port:PORT/port:
        PORT_LIST/port:name;
      }
    }
    leaf ip-prefix {
      type stypes:sonic-ip4-prefix;
    }
\end{verbatim}
\noindent \textbf{Agent}: \textit{Suggests configuration steps and confirms with user.}
\begin{verbatim}
  Assign IP addresses to Ethernet4 interfaces on 
  each device. You could use a subnet like 
  192.168.1.0/30 for each connection:
  L0 to S0: Assign 192.168.1.1/30 to L0 
  and 192.168.1.2/30 to S0.
  L0 to S1: ...
  L1 to S0: ...
  L1 to S1: ...
\end{verbatim}
\noindent \textbf{User}: \textit{Provides feedback on the suggestion.}
\begin{verbatim}
  I prefer you use 10.1.1.0/30 subnet instead.
\end{verbatim}
\noindent \textbf{Agent}: \textit{Suggests configuration steps and confirms with user.}
\begin{verbatim}
  Got it! Here’s the revised IP address 
  assignment for the Ethernet4 interfaces:
  L0 to S0: Assign 10.1.1.1/30 to L0 
  and 10.1.1.2/30 to S0.
  ...
\end{verbatim}

\noindent \textbf{User}: \textit{Confirms.}

\noindent \textbf{Agent}: \textit{Invokes the configuration generator.}

\noindent \textbf{Configuration Generator}:
\begin{verbatim}
  "device": L0,
  "action": "append",
  "path": "sonic-interface/sonic-interface/
    INTERFACE/INTERFACE_IPPREFIX_LIST",
  "value": {
    "name": "Ethernet4",
    "ip-prefix": "10.1.1.1/30"
  }
  ...
\end{verbatim}

\noindent \textbf{Agent}: \textit{Invokes the schema verifier.}

\noindent \textbf{Verifier}: 
\textit{Returns error message.}

\begin{verbatim}
  Must condition (current() = ../../INTERFACE_LIST
  [name=current()]/name) not satisfied
\end{verbatim}

\noindent \textbf{Agent}: \textit{Retrieves context about error from memory store.}

\noindent \textbf{Memory Store}:
\begin{verbatim}
  container sonic-interface {
    container INTERFACE {
      list INTERFACE_LIST {
        key "name";
        leaf name {
          type leafref {
            path /port:sonic-port/port:PORT/port:
             PORT_LIST/port:name;
    ...
\end{verbatim}

\noindent \textbf{Agent}: \textit{Invokes the configuration generator.}

\noindent \textbf{Configuration Generator}: 
\begin{verbatim}
  # Append the following to previous config
  "device": L0,
  "action": "append",
  "path": "sonic-interface/sonic-interface/
    INTERFACE/INTERFACE_LIST",
  "value": {
    "name": "Ethernet176"
  }
  ...
\end{verbatim}

\noindent \textbf{Agent}: \textit{Invokes the schema verifier.}

\noindent \textbf{Verifier}: \textit{Returns success message.}

\noindent \textbf{Agent}: \textit{Displays final configuration and confirms with user.}

\noindent \textbf{User}: \textit{Confirms.}

\noindent \textbf{Agent}: \textit{Deploy configuration to network. }

\newpage
\onecolumn

\section{Prompts}
\label{sec-appendix-b}

\newtcolorbox{titleEnv}{
    width=\linewidth,
    colframe=white,
    %valign=center, % or top
    before=\vspace{0pt}, % ensure no extra vertical space
    after=\vspace{0pt}   % ensure no added space after box
}

\begin{titleEnv}

\textbf{\textit{[System Prompt]: }}You are an AI assistant for network configuration. There are four devices in a 3-stage clos network, named `S0', `S1', `L0', `L1'. Please use these names exactly. A network operator wants to configure a network using natural language. You will be provided the operator's request in natural language. For context, you will be given several YANG modules that describe a schema for you to express the device configurations. Your job is generate ABNF changes in JSON format needed to perform the specific configuration given by the operator. Notice in the example that you are limited only to the operations ``append'' and ``remove'', and that the changes are given as paths based on the schema structure. 

You should list out the configurations for each device in the network. Your list should be expressed in Python syntax and clearly delineated so that it can be automatically extracted. Ensure that different devices are separate items in the list, although the entire configuration should be a single Python snippet for extraction. At the end, provide a brief explanation of your translation. 

\verb|**********************************************|

\textit{Example natural language query:} ``Hey, can you assign IP address 10.0.5.1/24 to the interface Ethernet8 of device S0? Replace the old IP address if there is one already at Ethernet8.''

\textit{Configuration:}
\begin{verbatim}
```python
[
    {{
        "device": "S0",
        "config": [
            {{
                "action": "remove",
                "path": ["sonic-interface:sonic-interface", "sonic-interface:INTERFACE", 
                         "INTERFACE_IPPREFIX_LIST"],
                "value": {{"name": "Ethernet8", "ip-prefix": "10.0.2.1/24"}}
            }},
            {{
                "action": "append",
                "path": ["sonic-interface:sonic-interface", "sonic-interface:INTERFACE", 
                         "INTERFACE_IPPREFIX_LIST"],
                "value": {{"name": "Ethernet8", "ip-prefix": "10.0.5.1/24"}}
            }}
        ]
    }}
]
'''
\end{verbatim}

\textit{Explanation:}
$\langle$Explain your answer here, and also list any assumptions about the network state here. In this example, you might assume that Ethernet8 already has an active IP address 10.0.2.1/24.$\rangle$

\bigskip

\textbf{\textit{[User Prompt]: }}Here is information about the current network state: \textbf{\{insert network state from retrieval here\}}

Here are the YANG modules that are relevant for this configuration:
\textbf{\{insert yang models from retrieval here\}}

\textit{Natural language query:} \textbf{\{insert user request here\}}

\textit{Configuration: }
\end{titleEnv}

\noindent\begin{minipage}{\textwidth}
\captionof{figure}{System and user prompts for the configuration generator.}\label{config-generator-prompt}

\end{minipage}

% ------------------------- SECOND PROMPT ------------------------- %

\begin{titleEnv}

\textbf{\textit{[System Prompt]: }}You are an assistant to a network operator trying to configure an SONIC clos network. The configuration code is given to you in a YANG ABNF data format, but it does not pass the YANG validator tests. Feedback error logs are given to you, as well as relevant YANG modules retrieved from a vector store containing all of the YANG modules, which specify the grammar to which the YANG configuration must adhere. Please correct the faulty YANG configuration and return the correct version clearly delineated as a standalone list within a single, complete Python snippet (see below). For automatic extraction purposes, the solution should be the only Python block in your output, as follows:

\begin{verbatim}
```python
[   <configuration>   ]
'''
\end{verbatim}

\bigskip

\textbf{\textit{[User Prompt]: }}\textit{Faulty YANG configuration: } \textbf{\{insert output from configuration generator here\}}

\textit{Error logs: } \textbf{\{insert verifier feedback here\}}

\textit{Retrieved YANG models: } \textbf{\{insert YANG retrieval results here\}}

\textit{Revised YANG Configuration: }

\end{titleEnv}

\noindent\begin{minipage}{\textwidth}
\captionof{figure}{System and user prompt for feedback interface tailored to verification.}\label{feedback-prompt}
\end{minipage}

\end{document}